\documentclass[aps,superscriptaddress,twocolumn,longbibliography]{revtex4-1}
\usepackage{times}
\usepackage{amsmath}
\usepackage{amsfonts}
\usepackage{amssymb}
\usepackage{mathtools}
\usepackage{mathrsfs}
\usepackage{color}
\usepackage[colorlinks=true,citecolor=blue,urlcolor=blue,linkcolor=blue,hyperindex]{hyperref}
\usepackage{wasysym}
\usepackage{bm}
\usepackage{cases}
\usepackage[version=4]{mhchem}
\usepackage{graphicx}
\usepackage{subfigure}
\allowdisplaybreaks

\begin{document}
\title{1-form symmetry and the selection rule of the plaquette valence bond solid phase on kagome lattice}
\author{Zijian Xiong}
\affiliation{Department of Physics, Chongqing University, Chongqing, 401331, China}
\affiliation{Chongqing Key Laboratory for Strongly Coupled Physics, Chongqing, 401331, China}
\affiliation{Department of Applied Physics, University of Tokyo, Tokyo 113-8656, Japan}

\begin{abstract}
We study the plaquette valence bond solid phase in a XXZ type spin-1/2 model in the kagome lattice. The low energy theory for this phase is a U(1) lattice gauge theory on the honeycomb lattice. We find that there is an emergent 1-form U(1) symmetry in low energy, and there is a mixed anomaly. We also show that this 1-form symmetry constraints the longitudinal dynamical structure factor and leads to the selection rule relating to the vanishing intensity along some high symmetry momentum paths (e.g. $\Gamma-M$ path). We point out that this emergent 1-form symmetry is robust against the translation symmetry preserving UV perturbation, thus the selection rule will also apply to the model which is obtained by perturbing the classical limit of our model.
\end{abstract}

\maketitle
\section{Introduction}
Symmetry is the most fundamental tool to study physics. It plays important roles from the classification of elementary particles to the understanding of phases and phase transitions. Conventionally,  the charged object under the symmetry is zero dimensional point like operator, and the corresponding symmetry transformation acts on the whole space. Recently, the concept of symmetry has been generalized \cite{Gaiotto2015,McGreevy2023}. Now, the charged object can be $p$ dimensional operator, and the symmetry transformation acts on the closed $(d-p)$ dimensional (i.e. codimension $p$) subspace of d dimensional space. Such symmetry is called a $p-$form symmetry, and the original symmetry corresponds to the 0-form symmetry.

Since the concept of the generalized symmetry has been proposed, the physics based on the original (0-form) symmetry has also been updated. For example, the Mermin-Wagner theorem is generalized to the higher form continuous symmetry case \cite{Gaiotto2015,Lake2018}, the 't Hooft anomaly \cite{thooft1980} involves the higher form symmetry is also studied \cite{Kobayashi2019prb,CMJian2021} and the symmetry protected topological phases are also generalized to include the higher form symmetry \cite{Yoshida2016prb,CMJian2021prb,Pace2023prb}. Among the studies of higher form symmetry, some notable findings are, i). the confined and deconfined phases can be distinguished by the unbroken/ broken 1-form symmetry in the sprit of the Landau-Ginzburg-Wilson symmetry breaking paradigm \cite{cordova2022snowmass}, ii). a subclass of topological ordered phases can be understood as higher form symmetry broken \cite{XGWen2019prb,pace2023}. Besides these, there are also many applications of higher form symmetry for constraining the phase and phase transition \cite{pace2023,Nahum2021prx,XCWu2021scipost}.

A basic example with the higher form symmetry is the Maxwell theory in three dimensional space, where there are two 1-form symmetries \cite{Gaiotto2015,DavidTong}. One of them is called electric 1-form, the symmetry transformation is associated with the two form conserved current $J^{e}\sim \mathord\star F$. Another is called magnetic 1-form, and it is associated with $J^{m}\sim F$ which is also conserved by Bianchi identity. The corresponding 1-form charged objects for these two symmetries are the Wilson line and 't Hooft line, respectively. 

In condensed matter physics, the low energy properties of certain systems are known to be well described by the emergent electromagnetic theory, or more generally, by emergent gauge theory. Some typical examples are quantum dimer model \cite{Fradkin1990mplb,Fradkin2013book}, quantum spin ice \cite{Hermele2004prb}, and quantum spin liquid \cite{Savary2017}. Specially, the low energy theories of the quantum spin ice and quantum dimer model on the bipartite lattice are both U(1) gauge theories \cite{Fradkin1990mplb}. The corresponding higher form symmetry in these bosonic lattice models have been identified \cite{pace2023,CMJian2021}. And some of the physical consequences of the higher form symmetry in these theories have also been studied, such as the properties of the higher form symmetry \cite{pace2023}, the stability of the gapless goldstone boson \cite{Hastings2005prb,Hofman2019,Hidaka2021prl}, 't Hooft anomaly \cite{Kobayashi2019prb} and spontaneous symmetry breaking \cite{Lake2018,pace2023}. In this paper, we study the selection rule for the dynamical structure factor from the 1-form U(1) symmetry. Concretely, we study the spin-1/2 model with strong Ising anisotropy on kagome lattice. And it should be noticed that some related models on kagome lattice have been realized in the Rydberg atom arrays \cite{Lukin2021pnas,Lukin2021science,Lukin2022prl}. 

This paper is organized as follows. In Sec.\ref{secmod}, we discuss the ground state manifold of the model and the quantum fluctuation. In Sec.\ref{gaugethe}, we map the low energy theory of the ground state manifold to a U(1) lattice gauge theory. We show that there is an emergent 1-form U(1) symmetry in the low energy and discuss its consequences. In Sec.\ref{secdsf}, we show that the 1-form symmetry constraints the longitudinal dynamical structure factor along some high symmetry momentum paths. Discussion about the high energy excitations are presented in Sec.\ref{secdis}, where a simple abelian Higgs model with charge 3 monopole is proposed. We show that there is no mixed anomaly in this abelian Higgs model. We also study the charge and domain wall excitations in this section. 

\section{Model}\label{secmod}
We first consider the spin-1/2 XXZ type model with Zeeman field on the kagome lattice around the Ising limit
\begin{equation}\label{ham}
H=-J_{xy}\sum_{\langle i,j\rangle}(S^{+}_{i}S^{-}_{j}+S^{-}_{i}S^{+}_{j})+J\sum_{\langle i,j\rangle}S^{z}_{i}S^{z}_{j}-h\sum_{i}S_{i}^{z},
\end{equation}
where $\langle i,j\rangle$ denotes the nearest-neighbor pair on the kagome lattice and $0<J_{xy}\ll J$. It is known that this model can be mapped to the extended Bose-Hubbard model with the mapping: $S_{i}^{-}\to b_{i}, S_{i}^{z}\to n_{i}-1/2$, where the up (down) spin is mapped to the presence (absence) of hardcore boson.

For the classical part, namely, the Ising model with Zeeman field
\begin{equation}\label{Isingm}
H^{0}=J\sum_{\langle i,j\rangle}S^{z}_{i}S^{z}_{j}-h\sum_{i}S_{i}^{z},
\end{equation}
due to the corner sharing nature of the kagome lattice, this model can be written as
\begin{equation}
H^{0}=\frac{J}{2}\sum_{p}[(\sum_{i\in p}S^{z}_{i})-\frac{h}{2J}]^2-C(h,J),
\end{equation}
where p denotes the smallest plaquette which consists of 3 spins, i.e. the up or down triangle $\triangle,\triangledown$ in the kagome lattice. $C(h,J)$ is a c number and only depends on h and J. The allowed values for $\sum_{i\in p}S^{z}_{i}$ are $-3/2, -1/2, 1/2, 3/2$. It is easy to check that when $-1<\frac{h}{2J}<0$, the Hamiltonian $H^{0}$ is minimized by $\sum_{i\in p}S_{i}^{z}=-1/2$ for every up and down triangle. This means that there are two down spins and one up spin in every triangle and corresponds to the so called 1/3 filling of the Bose-Hubbard model. Further, this condition also defines the classical ground state manifold which is extensively degenerate. Similarly, for $0<\frac{h}{2J}<1$, the Hamiltonian $H^{0}$ is minimized by $\sum_{i\in p}S_{i}^{z}=1/2$. In this situation, there are two up spins and one down spin in every triangle and corresponds to the 2/3 filling of the Bose-Hubbard model. 

\begin{figure}
	\centering
	\includegraphics[width=0.35\textwidth]{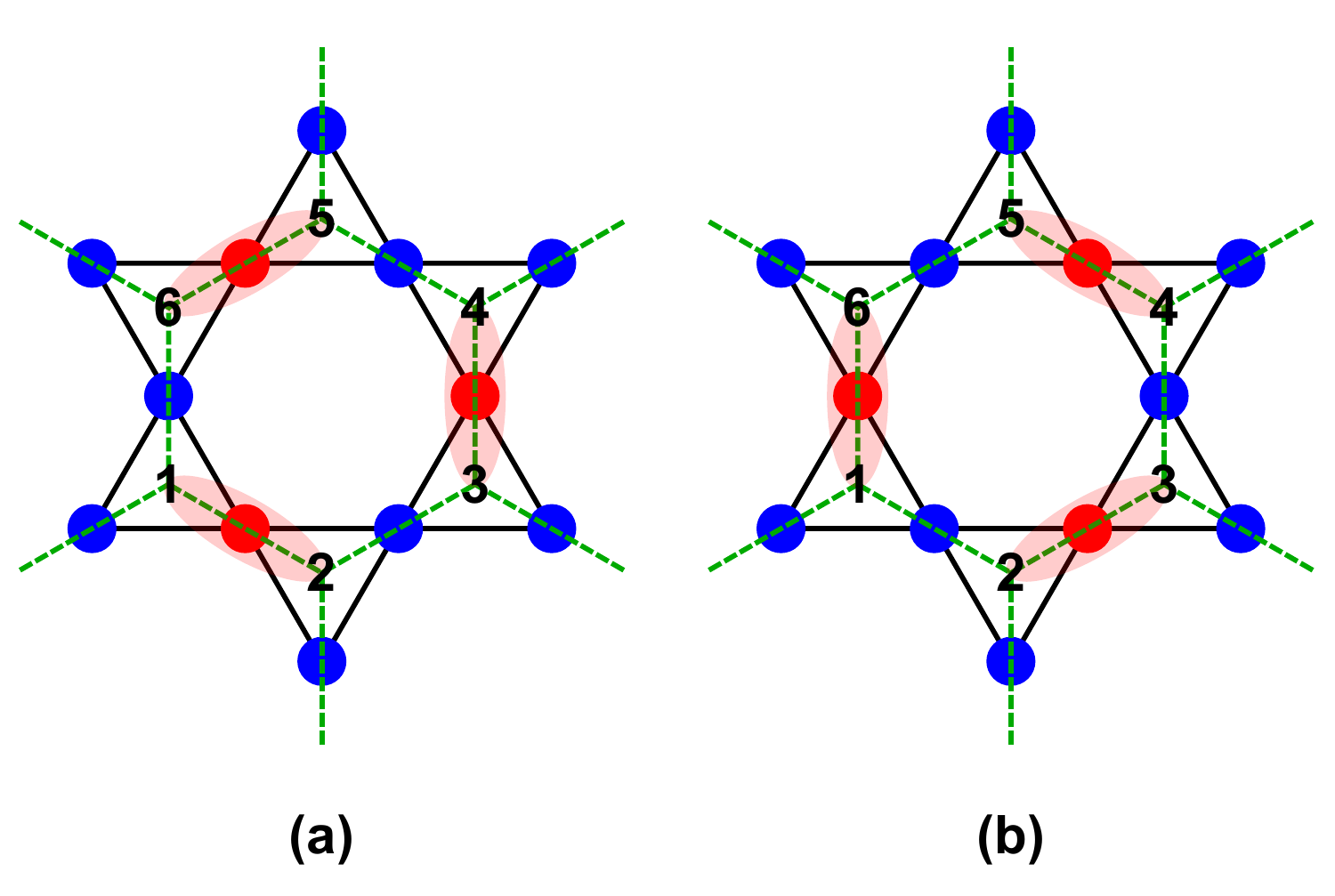}
	\caption{\label{lattice} Two ground state configurations of eq.(\ref{Isingm}). The red dots denote the up (down) spins for $-1<\frac{h}{2J}<0$ case where $\sum_{i\in p}S_{i}^{z}=-1/2$ for every triangle p (for $0<\frac{h}{2J}<1$, $\sum_{i\in p}S_{i}^{z}=1/2$), and the blue dots denote the down (up) spins. The red spin can be mapped to a dimer (red bar) in the honeycomb lattice (green dashed lattice). Configuration (a) and (b) can be transformed to each other by a ring exchange process.}
\end{figure}

In both $-1<\frac{h}{2J}<0$ and $0<\frac{h}{2J}<1$ cases, the model can be mapped to the dimer model on the honeycomb lattice \cite{Moessner2001prb,Isakov2006prl,Damle2006prl,Cabra2005prb}, where the up (down) spin corresponds to a dimer in the former (latter) case, see Fig.\ref{lattice}. It is clear that when $h=0$, the ground state manifold is complicated by mixing the $\sum_{i\in p}S_{i}^{z}=1/2$ and $\sum_{i\in p}S_{i}^{z}=-1/2$. And this case has been studied in Ref. \cite{nikoli2005prb,nikoli2005prb2,Isakov2006prl}. In the following, we will keep $h$ finite.

\begin{figure}
	\centering
	\includegraphics[width=0.28\textwidth]{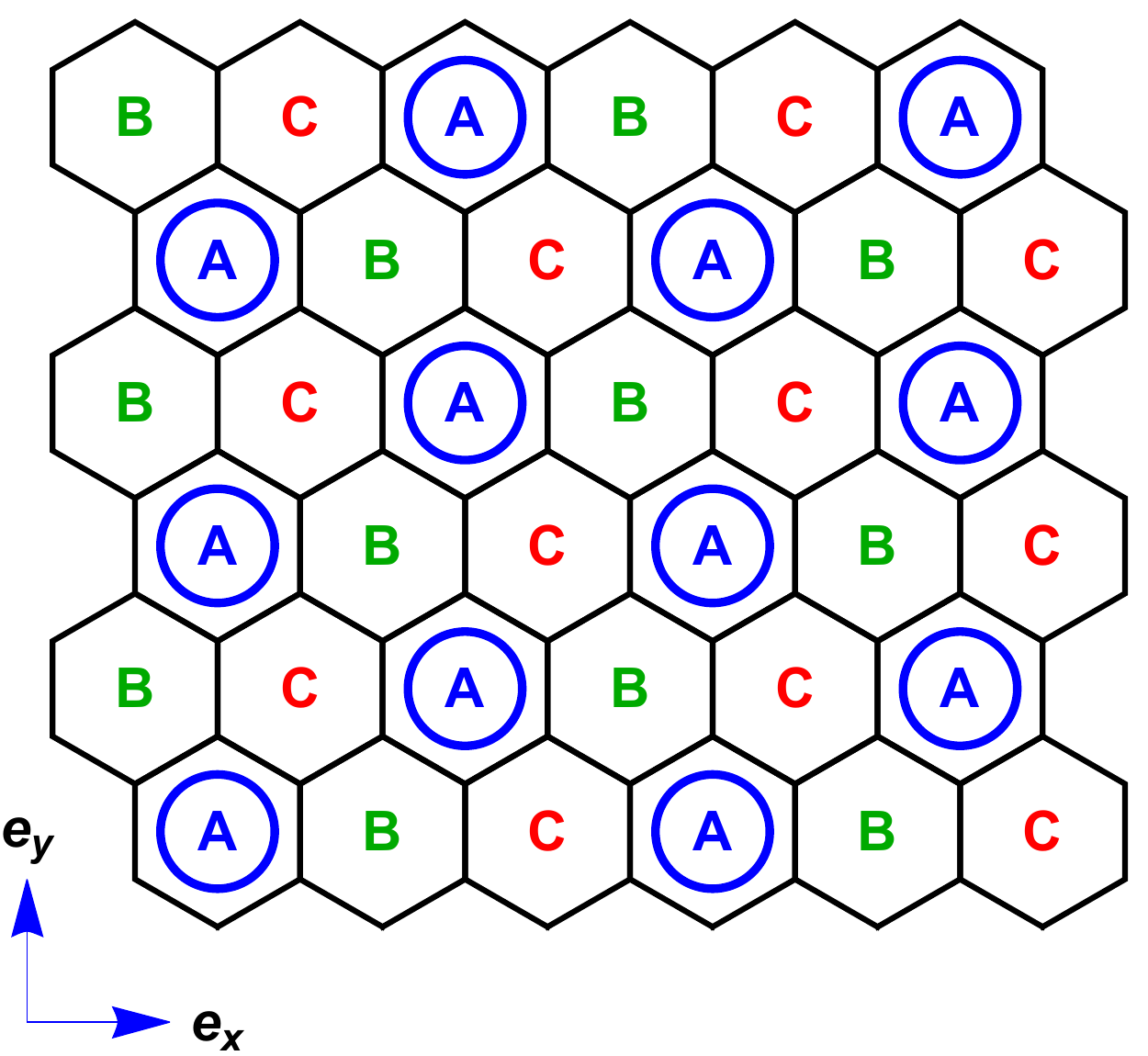}
	\caption{\label{pvbs} Plaquette valence bond solid order. A,B,C denote the three sublattices of the plaquette. The circles represent the flippable plaquettes.}
\end{figure}

Now we consider the effect of quantum fluctuation induced by $J_{xy}$ term in eq.(\ref{ham}). When $J_{xy}\ll J$, the leading effect is the quantum tunneling among the classical ground state configurations, such tunneling process can be captured by a six spin ring exchange effective Hamiltonian from the degenerate perturbation theory \cite{Cabra2005prb,XFZhang2013prl}
\begin{equation}
\begin{aligned}\label{ring}
H^{\rm{eff}}=-t_{ring}\sum_{\varhexagon}(&S^{+}_{\langle12\rangle}S^{-}_{\langle23\rangle}S^{+}_{\langle34\rangle}S^{-}_{\langle45\rangle}S^{+}_{\langle56\rangle}S^{-}_{\langle61\rangle}\\
&+S^{-}_{\langle12\rangle}S^{+}_{\langle23\rangle}S^{-}_{\langle34\rangle}S^{+}_{\langle45\rangle}S^{-}_{\langle56\rangle}S^{+}_{\langle61\rangle}),
\end{aligned}
\end{equation}
where $t_{ring}=12J_{xy}^{3}/V^{2}$, spin in the kagome site is represented by a bond variable $S^{\alpha}_{\langle rt\rangle}$ in the honeycomb lattice, see Fig.\ref{lattice}. And $1,2,3,4,5,6$ label the spins in the hexagonal plaquette $\varhexagon$ sequentially as in Fig.\ref{lattice}. The summation is over all the hexagonal plaquette. It is clear that the alternating up and down spins in a hexagonal plaquette will be flipped by this effective Hamiltonian. For example, the configurations (a) and (b) in Fig.\ref{lattice} can be flipped to each other. Since these flippable plaquettes have lower energy than the unflippable one under the tunneling process, and the flippable plaquettes repel each other, thus the ground state is expected to be a three fold degenerate plaquette valence bond solid phase (PVBS) with broken translation symmetry \cite{MoessnerHexagon2001prb,Mosseri2017prb}, see Fig.\ref{pvbs}. The flippable plaquettes occupy one of the three sublattices of the plaquette. Actually, the corresponding Bose-Hubbard model has been studied numerically, and this PVBS phase has been confirmed \cite{Isakov2006prl}. The phase diagram consists of a PVBS phase for $J_{xy}\ll J$ and a superfluid phase for $J_{xy}\gg J$ in both 1/3 filling and 2/3 filling cases. However, $h=0$ case is special, where the ground state configurations satisfies $\sum_{i\in p}S_{i}^{z}=\pm1/2$. And there is only a superfluid phase in the phase diagram of $h=0$ case. In this paper, we are interested in the PVBS phase in $-1<\frac{h}{2J}<0$, i.e. the 1/3 filling case.

\section{Low energy effective theory}\label{gaugethe}
As we shown in the previous section, the ground state configuration of eq.(\ref{ham}) for $-1<\frac{h}{2J}<0$ can be mapped to the dimer configuration. Hence, the low energy effective theory of the original model is a dimer model. We first introduce the rotor representation for the spin \cite{Fradkin1990mplb,Fradkin2013book}
\begin{equation}
\begin{aligned}
S^{z}_{\langle rt\rangle}=&\,n_{\langle rt\rangle}-\frac{1}{2},\\
S^{\pm}_{\langle rt\rangle}=&\,e^{\pm i\phi_{\langle rt\rangle}},
\end{aligned}
\end{equation}
$n_{\langle rt\rangle}=1 (0)$ for up (down) spin, and it represents the dimer number on the bond $\langle rt\rangle$. The canonical commutation relation is nonzero only when n and $\phi$ live in the same bond, $[\phi_{\langle rt\rangle},\,n_{\langle rt\rangle}]=i$. Moreover, both n and $\phi$ are bond scalar \cite{nikoli2005prb,nikoli2005prb2}: $f_{\langle rt\rangle}=f_{\langle tr\rangle}$. 

Usually, $n_{\langle rt\rangle}$ can be relaxed to take all integer values by introducing a Lagrange multiplier term $\frac{1}{2g}\sum_{\langle rt\rangle}[(n_{\langle rt\rangle}-\frac{1}{2})^2-\frac{1}{4}]$. The original values can be recovered by taking the limit $g\to 0$. In the honeycomb lattice, the total number of dimers $\sum_{\langle rt\rangle}n_{\langle rt\rangle}$ equals the total number of plaquette under the hardcore dimer constraint (every site in the honeycomb lattice has one and only one dimer) \cite{Mosseri2017prb,Mosseri2015prl}. Suppose the lattice has $L_{x}$ plaquettes along $e_{x}$ direction and $L_{y}$ plaquettes along the $e_{y}$ direction (see Fig.\ref{pvbs}), then $\sum_{\langle rt\rangle}n_{\langle rt\rangle}=L_{x}*L_{y}$.  As $n_{\langle rt\rangle}$ takes integer value, $\phi_{\langle rt\rangle}$ is an angular variable and takes value in $[0,2\pi)$.

With the rotor representation, the ground state condition can be written as the hardcore dimer constraint on every site of the honeycomb lattice
\begin{equation}
\begin{aligned}\label{hd1}
&\sum_{i\in p}S_{i}^{z}= -\frac{1}{2},\\
\to& \sum_{t\in r}n_{\langle rt\rangle}=1,
\end{aligned}
\end{equation}
where every site has one and only one dimer. The ring exchange term eq.(\ref{ring}) can be written as
\begin{equation}
\begin{aligned}\label{ring2}
&S^{+}_{\langle12\rangle}S^{-}_{\langle23\rangle}S^{+}_{\langle34\rangle}S^{-}_{\langle45\rangle}S^{+}_{\langle56\rangle}S^{-}_{\langle61\rangle}\\
&+S^{-}_{\langle12\rangle}S^{+}_{\langle23\rangle}S^{-}_{\langle34\rangle}S^{+}_{\langle45\rangle}S^{-}_{\langle56\rangle}S^{+}_{\langle61\rangle}\\
=&2\cos(\phi_{\langle 12\rangle}-\phi_{\langle 23\rangle}+\phi_{\langle 34\rangle}-\phi_{\langle 45\rangle}+\phi_{\langle 56\rangle}-\phi_{\langle 61\rangle}).
\end{aligned}
\end{equation}

Since the honeycomb lattice is a bipartite lattice, we define a function $\eta_{r}=1$ for the site $r$ which belongs to the sublattice $\mathcal{A}$ and $\eta_{r}=-1$ for sublattice $\mathcal{B}$. Then we can define the lattice fields 
\begin{equation}
\begin{aligned}
E_{rt}=&\eta_{r}n_{\langle rt\rangle},\\
a_{rt}=&\eta_{r}\phi_{\langle rt\rangle},
\end{aligned}
\end{equation}
they are bond vectors \cite{nikoli2005prb,nikoli2005prb2}: $f_{rt}=-f_{tr}$. And the commutation relation is $[a_{rt}, E_{rt}]=i$. Two useful relations can be derived from the commutation relation
\begin{equation}
\begin{aligned}\label{usefuleq}
[e^{i\omega\,a_{rt}, E_{rt}}]=&-\omega e^{i\omega\,a_{rt}},\\
[a_{rt}, e^{i\omega\,E_{rt}}]=&-\omega e^{i\omega\,E_{rt}},
\end{aligned}
\end{equation}
where $\omega$ is a c number.

Now, the hardcore dimer constraint eq.(\ref{hd1}) can be written as the Gauss's law. The lattice divergence is defined as
\begin{equation}
(\rm{div}\, \mathnormal{E})_{r}\equiv\sum_{t\in r}\mathnormal{E}_{rt}=\eta_r,
\end{equation}
this implies that there are static background charge +1(-1) on the sites of sublattice $\mathcal{A}$ ($\mathcal{B}$). 

The ring exchange term eq.(\ref{ring2}) can be expressed as a lattice curl
\begin{equation}
\begin{aligned}\label{latcurl}
&\cos(\phi_{\langle 12\rangle}-\phi_{\langle 23\rangle}+\phi_{\langle 34\rangle}-\phi_{\langle 45\rangle}+\phi_{\langle 56\rangle}-\phi_{\langle 61\rangle}) \\
=&\cos(a_{12}+a_{23}+a_{34}+a_{45}+a_{56}+a_{61})\\ 
\equiv&\cos[(\rm{curl}\,\textit{a})_1],
\end{aligned}
\end{equation}

\begin{figure}
	\centering
	\includegraphics[width=0.15\textwidth]{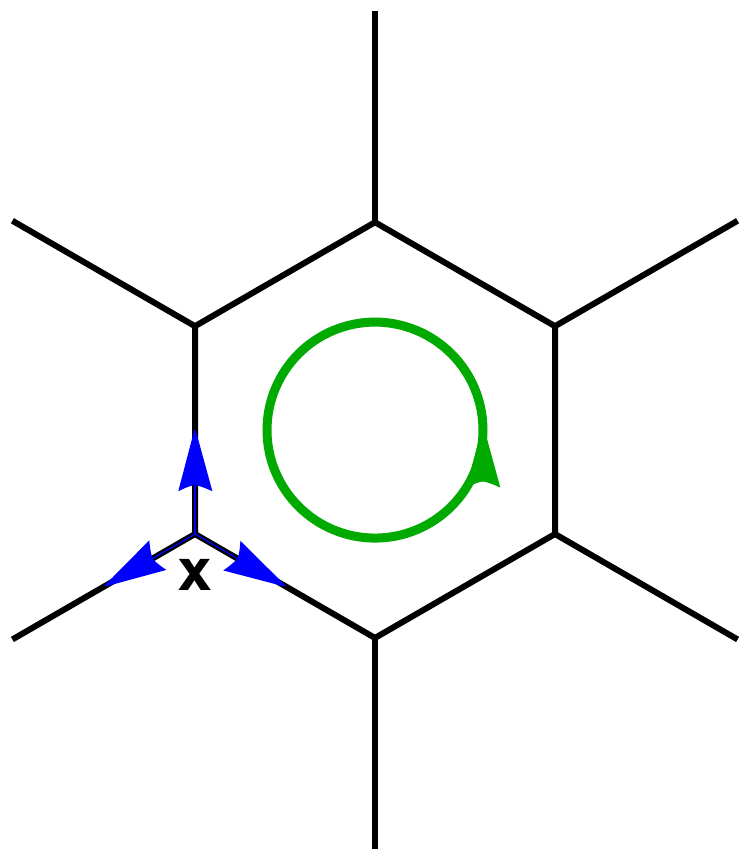}
	\caption{\label{divcurl} A simple way to see $[(\rm{div}\,\mathnormal{E})_{r}, (\rm{curl\,\textit{a}})_{r}]=0$. The blue arrows denote the E fields in $(\rm{div}\,\mathnormal{E})$. The green counterclockwise circle denote the directions of $a$ fields in $(\rm{curl\,\textit{a}})$. When $a$ and $E$ are on the same bond, $[a,E]=\pm i$ if $a$ and $E$ are parallel (antiparallel).}
\end{figure}

Finally, the low energy effective theory takes a compact quantum electrodynamics (cQED) form
\begin{equation}\label{cQED}
H^{\rm{cQED}}=\frac{1}{2g}[(\sum_{rt}E_{rt}^2)-L_{x}L_{y}]-2t_{ring}\sum_{r\in \mathcal{A}}\cos[(\rm{curl}\,\textit{a})_r],
\end{equation}
where $rt$ denotes the nearest neighbor bond on the honeycomb lattice. If $a$ and E fields are on the same bond, we have $[a,E]=\pm i$ if $a$ and E are parallel (antiparallel). Now consider the case in Fig.\ref{divcurl}, the counterclockwise circle denotes the directions of $a$ fields in $(\rm{curl\,\textit{a}})$, and the blue arrows denote the E fields in $(\rm{div}\,\mathnormal{E})$. Then, it is easy to find $[(\rm{div}\,\mathnormal{E})_{r}, (\rm{curl\,\textit{a}})_{t}]=0$ for any r,t sites, since there is one parallel $a$-$E$ pair and one antiparallel pair. 
And $[(\rm{div}\,\mathnormal{E})_{r}, \textit{H}^{\rm{cQED}}]=0$.

The gauge symmetry is generated by $U=e^{i\sum_{r}\theta_{r}(\rm{div}\, \mathnormal{E})_{r}}$, where $\theta_{r}$ is a site dependent c number. So $H^{\rm{cQED}}$ is gauge invariant. Using eq.(\ref{usefuleq}), we can find 
\begin{equation}
U\,a_{rt}\,U^{\dagger}=a_{rt}+\theta_{r}-\theta_{t},
\end{equation}
which shows that $a$ field transforms as the vector potential. The gauge symmetry here actually origins from the local rotation symmetry $e^{i\alpha_{p}\sum_{i\in p}S^{z}_{p}}$ in the ground state manifold. This symmetry is imposed by the large energy scale of $J$. Thus the low energy effective theory is required to be gauge invariant, equivalently, the ground state condition $\sum_{i\in p}S^{z}_{p}=-1/2$ should be satisfied for every $p$. And the simplest gauge invariant dynamics is induced by the 6 sites loop operator- the lattice curl term eq.(\ref{latcurl}). The same term can also be generated in low energy by other quantum fluctuation rather than the $J_{xy}$ term, for example, the transverse field term \cite{Moessner2001prb,GChen2019prr}. From this gauge invariant principle, it is easy to know that the higher order dynamics is also the (contractible) loop operator but involves more sites.

Further, it should be noticed that the low energy effective theory above breaks the charge conjugation symmetry ($E\to -E, a\to -a$) implicitly. This can be seen by adding a term $\frac{J}{2}\sum_{r}[(\rm{div}\,\mathnormal{E})_{r}-\eta_r]^2$ to impose the Gauss's law explicitly.

\subsection{1- form U(1) global symmetry}
\begin{figure}
	\centering
	\includegraphics[width=0.48\textwidth]{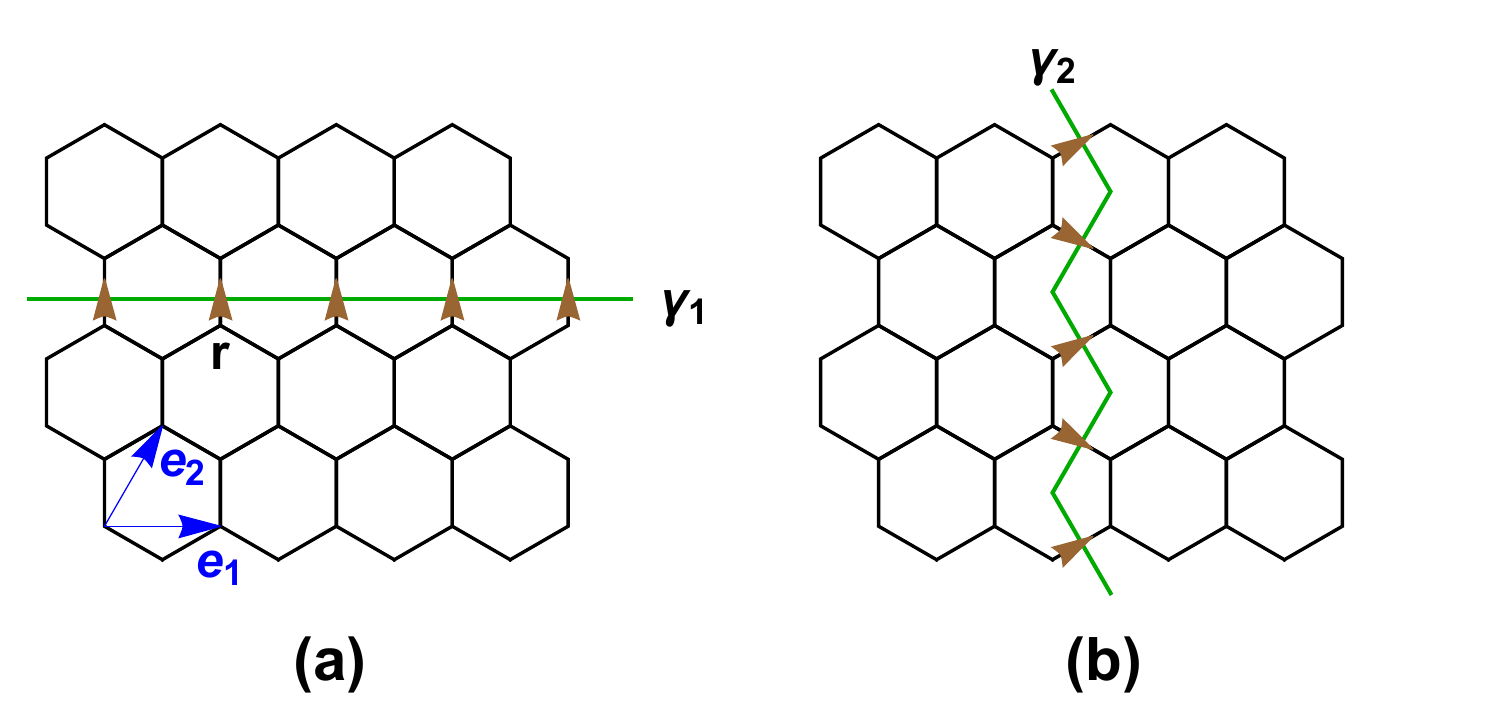}
	\caption{\label{winding} Two noncontractible loops $\gamma_1$ and $\gamma_2$ under the periodic boundary condition. The 't Hooft loop operator is constructed by summing over all $E$ fields which are passing through by the noncontractible loop and the fields are orthogonal to the loop as denoted by the arrows. The primitive translation vectors are $e_{1}$ and $e_{2}$. }
\end{figure}

Inspired by Fig.\ref{divcurl}, we can construct more conserved terms. For example, one can find
\begin{equation}
[(\rm{curl}\,\textit{a})_{r'}, \sum_{n=0}^{\mathnormal{L}_{\mathnormal{x}}-1}\mathnormal{E}_{\mathnormal{y}}(\mathnormal{r_x+n, r_y})]=0,
\end{equation}
as shown in Fig.\ref{winding} (a), where $E_{y}(r_x+n, r_y)$ is the bond vector from site $(r_x+n, r_y)$ to site $(r_x+n, r_y+b)$. $b$ is the nearest neighbor bond length. The lattice constant of the honeycomb lattice is set to be 1. $r=(r_{x},r_{y})$ is a site near to the loop $\gamma_1$, and the bond from site $r$ to site $r+b \vec{e}_{y}$ is passed through by loop $\gamma_1$. This is actually a 't Hooft loop operator \cite{Fradkin2013book,rothe}
\begin{equation}
\widetilde{W}_{\gamma_1}=\sum_{n=0}^{L_{x}-1}E_{y}(r_x+n, r_y),
\end{equation}
its eigenvalue is usually called winding number. Another 't Hooft loop operator $\widetilde{W}_{\gamma_2}$ can be constructed as shown in Fig.\ref{winding} (b). 

\begin{figure}
	\centering
	\includegraphics[width=0.25\textwidth]{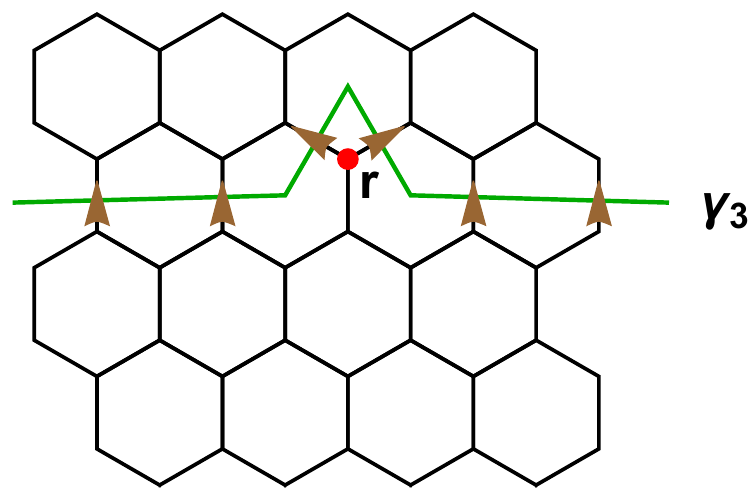}
	\caption{\label{deform} Deform the 't Hooft loop operator by adding a lattice divergence (red dot) term.}
\end{figure}

Further, one can find other different conserved 't Hooft loop operators. But all of them can be deformed to the above two. For example, we show a specific deformation in Fig.\ref{deform}: $\widetilde{W}_{\gamma_1}$ can be deformed to $\widetilde{W}_{\gamma_3}$ by adding a lattice divergence term $(\rm{div}\,\mathnormal{E})_{r}$. Since $(\rm{div}\,\mathnormal{E})_{r}$ commutes with $\textit{H}^{\rm{cQED}}$, so $\widetilde{W}_{\gamma_3}$ is also a conserved term. Using this strategy, one can check that $\widetilde{W}_{\gamma_1}$ can not be transformed to $\widetilde{W}_{\gamma_2}$. 

With this strategy, we can also define the topological loop operator by subtracting all the lattice divergence terms in the "bulk" from 't Hooft loop operator $\widetilde{W}_{\gamma_1}$, here the "bulk" can be chosen as the area below $\gamma_1$. In this definition, one can check that the topological loop operator will be invariant under such deformation. Now $\gamma_1$ can be regarded as a boundary, and the $E$ fields on it are all along the normal direction and pointing "outside". 

\begin{figure}
	\centering
	\includegraphics[width=0.48\textwidth]{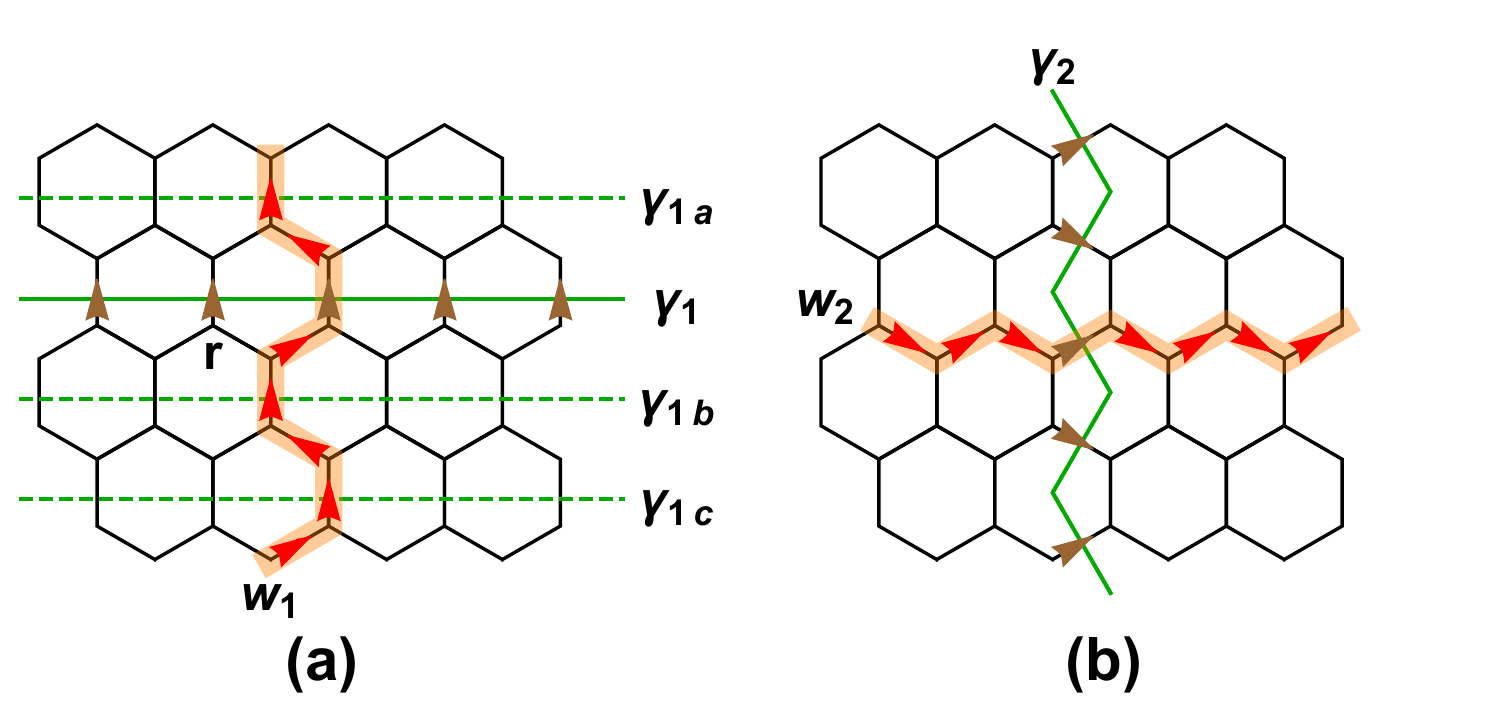}
	\caption{\label{wilson} Wilson loop operators on two noncontractible loops (orange paths) (a) $\rm{w}_{1}$ and (b) $\rm{w}_{2}$. The red arrows denote the $a$ fields in the Wilson loop. The green dashed lines denote equivalent loops for 't Hooft operator.}
\end{figure}

It is obvious that we can also construct more conserved terms by summing over different 't Hooft loops. For example, sum over 't Hooft operators on $\gamma_{1a}, \gamma_{1}, \gamma_{1b},\gamma_{1c},\cdots $, see Fig.\ref{wilson} (a). This is related to the fractonic symmetry \cite{MengCheng2019prb}.

Now we define the Wilson loop operator on a noncontractible loop $\rm{w}_{1}$, see Fig.\ref{wilson} (a)
\begin{equation}
W_{\rm{w}_1}=e^{i\sum_{rt\in \rm{w}_{1}}a_{rt}},
\end{equation}
$a_{rt}$ are along the path $\rm{w}_{1}$ as denoted by the red arrows. 

Recall the eq.(\ref{usefuleq}), one can find
\begin{equation}
\begin{aligned}\label{wilthoo}
[W_{\rm{w}_1}, \widetilde{W}_{\gamma_1}]=&-W_{\rm{w}_1},\\
[W_{\rm{w}_1}^{\dagger}, \widetilde{W}_{\gamma_1}]=&W_{\rm{w}_1}^{\dagger}.
\end{aligned}
\end{equation}
Suppose $|\widetilde{w}\rangle$ is the eigenstate of the 't Hooft loop $\widetilde{W}_{\rm{w}_1}$, and the eigenvalue is $\widetilde{w}$, then 
\begin{equation}
\begin{aligned}
\widetilde{W}_{\gamma_1}W_{\rm{w}_1}|\widetilde{w}\rangle=&(\widetilde{w}+1)W_{\rm{w}_1}|\widetilde{w}\rangle,\\
\widetilde{W}_{\gamma_1}W_{\rm{w}_1}^{\dagger}|\widetilde{w}\rangle=&(\widetilde{w}-1)W_{\rm{w}_1}^{\dagger}|\widetilde{w}\rangle,
\end{aligned}
\end{equation}
so the Wilson operators $W_{\rm{w}_1}$ and $W_{\rm{w}_1}^{\dagger}$ are the ladder operators for the 't Hooft loop $\widetilde{W}_{\gamma_1}$. Another Wilson loop $W_{\rm{w}_2}$ (see Fig.\ref{wilson} (b)) has the same properties. 

 Based on eq.(\ref{wilthoo}), one can find 
 \begin{equation}
 e^{i\alpha\widetilde{W}_{\gamma_1}}W_{\rm{w}_1} e^{-i\alpha\widetilde{W}_{\gamma_1}}= e^{i\alpha}W_{\rm{w}_1},
 \end{equation}
 this is indeed the definition of the 1-form symmetry \cite{Gaiotto2015,McGreevy2023}, where $e^{i\alpha\widetilde{W}_{\gamma_1}}$ is the 1-form symmetry operator, and the 't Hooft loop $\widetilde{W}_{\gamma_1}$ is the 1-form charge. This charge is carried by the Wilson loop $W_{\rm{w}_1}$. $a$ field which is passed through by $\gamma_1$ will transforms as
 \begin{equation}
 e^{i\alpha\widetilde{W}_{\gamma_1}}a_{xy} e^{-i\alpha\widetilde{W}_{\gamma_1}}=a_{xy}+\alpha,
\end{equation}
since $(\rm{curl\, \textit{a}})$ should be invariant under the transformation, thus $(\rm{curl\, \alpha})=0$. In other words, $\alpha$ is a flat 1-form connection. From above equation, one can find that the 1-form symmetry transformation shifts the $a$ field by a flat 1-form connection.

It is pointed out that the emergent higher form symmetry is exact in Ref.\cite{Hastings2005prb,pace2023}, in the sense that the emergent higher form symmetry are robust against any local UV perturbations which preserve the translation symmetry. Actually, the discussion of the gauge invariant principle in Sec.\ref{gaugethe} also implies this robustness, since the quantum fluctuation in UV will act as contractible loop operator in the low energy states. And we know that among the gauge invariant operators, only the noncontractible Wilson loop operator is charged under the higher form symmetry. But such charged operators are nonlocal, thus they will not be generated by the local UV perturbations.

\subsection{Adiabatic flux insertion}
Now we consider the physical consequence from the 1-form symmetry by the adiabatic flux insertion process \cite{Oshikawa2000prl}. Consider a translation invariant system with finite many body gap at time t=0, we insert a flux couples to the 1-form U(1) global symmetry adiabatically. Suppose the many body gap does not close and the translation symmetry is unchanged during the flux insertion process. Denote the ground state of $H(t=0)$ as $|\psi_0\rangle$. It is also the eigenstate of momentum, say $T_{1}|\psi_0\rangle=e^{ip_0}|\psi_0\rangle$, where $T_1$ is the translation operator along $e_1$ direction (see Fig.\ref{winding} (a)). After we insert an unit flux adiabatically, the Hamiltonian becomes $H(\phi)$ and the initial state evolves to $|\psi'_0\rangle$ with the same momentum $p_0$. Since the unit flux is equivalent to zero flux, the effect of the flux can be eliminated by a large gauge transformation $U_{L}H(\phi)U^{-1}_{L}=H(t=0)$. Then the final state $|\psi'_0\rangle$ transforms to $U_{L}|\psi'_0\rangle$, which should be one of the ground state since we assume the many body gap does not close during the flux insertion.

Concretely, the translation operator along the $e_{1}$ direction acts as: $T_{1}^{\dagger}E_{y}(r_{x},r_{y})T_{1}=E_{y}(r_{x}-1,r_{y})$. And the large gauge transformation is given by $U_{L}=\rm{exp}[\mathnormal{-\frac{2\pi i}{L_x}\sum_{n=0}^{L_x-1}(r_x+n)E_{y}(r_{x}+n,r_{y})}]$. One can find
\begin{equation}
T_{1}^{\dagger}U_{L}T_{1}=U_L \rm{exp}[2\pi i \mathnormal{E_{y}(r_x,r_y)}]exp[\mathnormal{-\frac{2\pi i}{L_x}\widetilde{W}_{\gamma_1}}],
\end{equation}
since $E$ field takes integer value, so $\rm{exp}[2\pi i \mathnormal{E_{y}(r_x,r_y)}]=1$. Then we have
\begin{equation}
T_{1}U_{L}|\psi'_0\rangle=\rm{exp}[\mathnormal{\frac{2\pi i}{L_x}\widetilde{W}_{\gamma_1}}]\mathnormal{e^{ip_0}U_L}|\psi'_0\rangle,
\end{equation}
which shows $U_{L}|\psi'_0\rangle$ is indeed a eigenstate of momentum. If the filling of the 1-form charge $\widetilde{W}_{\gamma_1}/L_x$ is not an integer, then the momentum of $U_L|\psi'_0\rangle$ will different from $p_0$, which implies the ground state is degenerate. This is actually a Lieb-Schultz-Mattis \cite{LSM1961} type constraint for ground state of the translation invariant system with 1-form U(1) global symmetry \cite{Kobayashi2019prb}, which prohibits the nondegenerate symmetric ground state if the filling of the 1-form charge is not an integer. And it is known that the continuous 1-form global symmetry can not be spontaneously broken in two spatial dimension \cite{Gaiotto2015,Lake2018}. Thus, for a translation invariant system with 1-form U(1) global symmetry, if the filling of the 1-form charge is not an integer, the ground state can be i). topological ordered, ii). translation symmetry breaking or iii). gapless. This constraint can be also regarded as a mixed 't Hooft anomaly between the 1-form symmetry and translation symmetry \cite{Kobayashi2019prb,McGreevy2023}. It should be noticed that the original spin model eq.(\ref{ham}) does not constrained by any Lieb-Schultz-Mattis theorem, since there is a trivial paramagnet phase for large Zeeman field. In some literatures, this emergent Lieb-Schultz-Mattis constraint in low energy is called emergent anomaly \cite{Metlitski2018prb}.

For PVBS phase in Fig.\ref{pvbs}, the filling of the 1-form charge is 1/3, and it is a translation symmetry breaking phase which is consistent with above Lieb-Schultz-Mattis type constraint. 

\subsection{Duality transformation and height representation}
\begin{figure}
	\centering
	\includegraphics[width=0.48\textwidth]{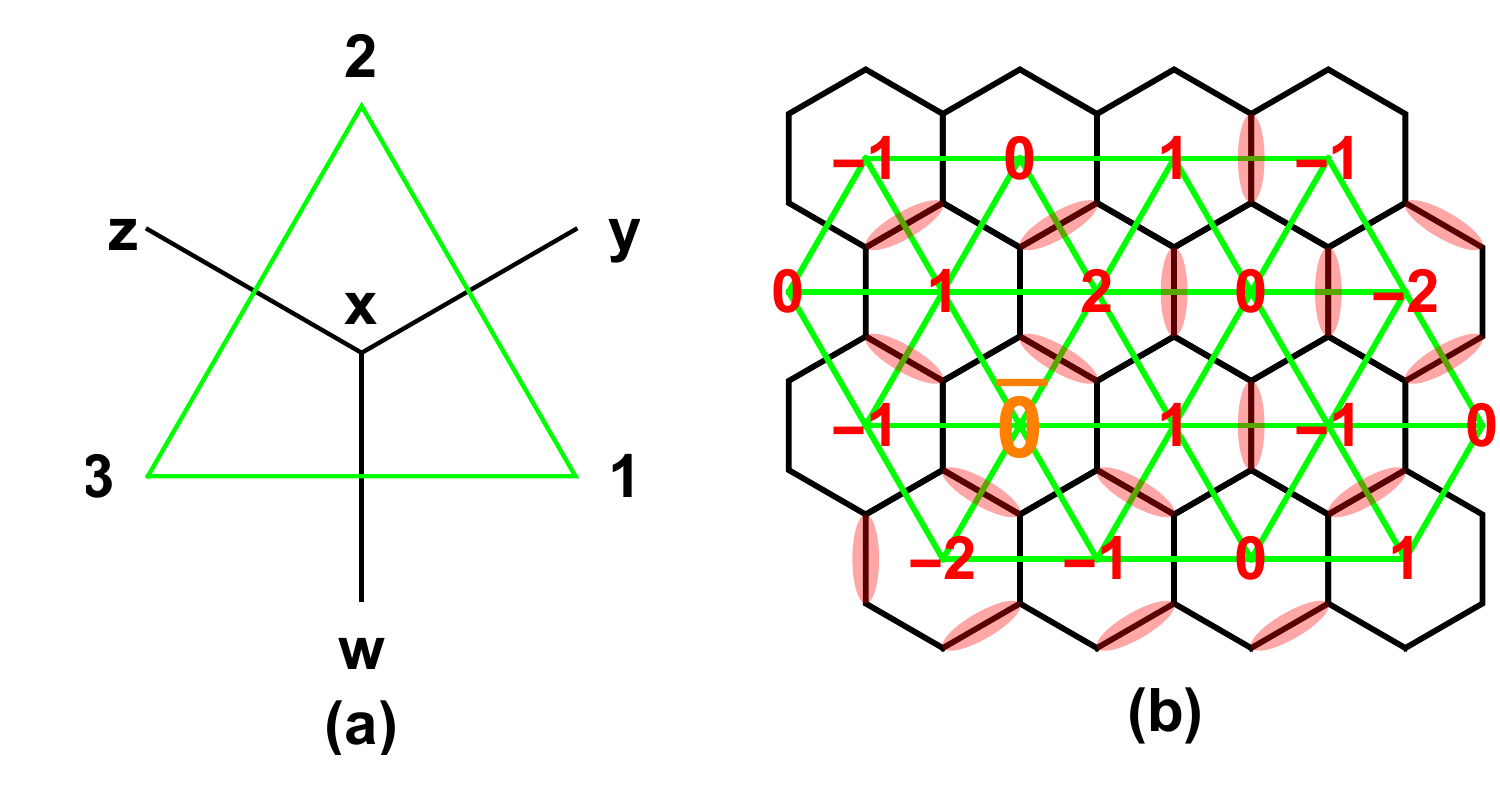}
	\caption{\label{dual} (a) $E$ field can be expressed by the difference of $h$ fields on the sites of triangle lattice and $B$ field on the link of triangle lattice. For example, $E_{xy}=h_{2}-h_{1}+B_{12}$. (b) Dimer configuration on the honeycomb lattice is equivalent to the height field configuration on the triangle lattice. $\bar{0}$ is the reference site.}
\end{figure}

Besides the gauge theory we used above, height field \cite{Henley1997,Moessner2001prb,Fradkin2004prb} is also usually used to describe the dimer model. Here we will show that the height field description is actually a dual theory for the gauge theory. It is known that the dual lattice of the honeycomb lattice is the triangle lattice \cite{Moessner2001prb}, see Fig.\ref{dual} (b). In the first step, we introduce the field lives on the dual site and dual link
\begin{equation}\label{duality}
E_{xy}=h_{left}-h_{right}+B_{\vec{z}\times \widehat{xy}},
\end{equation}
where $h$ is defined on the site of the triangle lattice, and $B$ is defined on the link of the triangle lattice. For example, consider $E_{xy}$ in Fig.\ref{dual} (a), $h_{left}=h_{2}$, $h_{right}=h_{1}$ and $B_{\vec{z}\times \widehat{xy}}=B_{12}$. So the lattice divergence can be written as
\begin{equation}
\begin{aligned}
(\rm{div}\,\mathnormal{E})_{x}=&E_{xy}+E_{xz}+E_{xw}\\
=&h_{2}-h_{1}+B_{12}+h_{3}-h_{2}+B_{23}\\
&+h_{1}-h_{3}+B_{31}\\
=&B_{12}+B_{23}+B_{31}.
\end{aligned}
\end{equation}
We see that the lattice divergence is transformed to lattice curl by transformation eq.(\ref{duality}). 

For the ground state manifold, $B_{12}+B_{23}+B_{31}=\eta_{x}$. We can define $B_{mn}=3\eta_{x}$ if $B_{mn}$ crosses a dimer, and $B_{mn}=-\eta_{x}$ if $B_{mn}$ does not cross a dimer. Then $h$ field is determined by eq.(\ref{duality}). We can summarize a rule for $h$: go around the up triangle clockwise in the triangle lattice, $h$ will increases by 2 if one crosses a dimer, and decreases by 1 if one does not meet a dimer. With this rule for $h$ field,  if a dimer configuration is given and choose a reference site where $h=0$, then this dimer configuration can be completely characterized by the $h$ field configuration, for example, see Fig.\ref{dual} (b). Actually, $h$ field is the so called height field \cite{Henley1997,Moessner2001prb,Fradkin2004prb}.

The monopole effect in the compact QED eq.(\ref{cQED}) can be studied by path integral \cite{Fradkin2013book}, after coarse graining the height field and taking the dilute monopole gas approximation, one can find a sine-Gordon theory \cite{Sachdev1990prb,Cenke2011prb}
\begin{equation}\label{sing}
S_{sg}=\int d^{2}xd\tau[(\partial_{\mu}\Phi)^2-\gamma\cos{(6\pi\Phi)}],
\end{equation}
where $\Phi$ is the coarse-grained height field, the $\cos$ term describes the triple monopole. And the three independent minima of the potential describe the threefold degenerate PVBS phase. This phase is a confined phase \cite{Polyakov1977}, equivalently, the 1-form U(1) symmetric phase \cite{cordova2022snowmass}.

\section{Selection rule from 1-form symmetry}\label{secdsf}
Now we consider the longitudinal dynamical structure factor of this model eq.(\ref{ham}), which is studied by quantum Monte-Carlo recently \cite{liu2023}
\begin{equation}
S^{zz}(\mathbf{Q},\omega)=\int \frac{dt}{2\pi}e^{i\omega t}\langle S^{z}_{\mathbf{Q}}(t)S^{z}_{-\mathbf{Q}}(0)\rangle,
\end{equation}
where $S^{z}_{\mathbf{Q}}=\sum_{\alpha}S^{z}_{\mathbf{Q},\alpha}$, and $S^{z}_{\mathbf{Q},\alpha}=\frac{1}{N}\sum_{i}e^{-i\mathbf{Q}\cdot \mathbf{r}_{i,\alpha}}S_{i,\alpha}^{z}$, $\mathbf{r}_{i,\alpha}=\mathbf{R}_{i}+\rho_{\alpha}$. $i$ labels the unitcell and $\alpha=1,2,3$ labels the sublattice of the kagome lattice, see Fig.\ref{dsf}. $\rho_{\alpha}$ is the position of the $\alpha$ spin in a unitcell, which are read as $\rho_1=\frac{1}{2\sqrt{3}}(-\frac{\sqrt{3}}{2},-\frac{1}{2})$, $\rho_2=\frac{1}{2\sqrt{3}}(\frac{\sqrt{3}}{2},-\frac{1}{2})$ and $\rho_3=\frac{1}{2\sqrt{3}}(0,1)$. 

\subsection{$q_{x}=0$ path}
\begin{figure}
	\centering
	\includegraphics[width=0.44\textwidth]{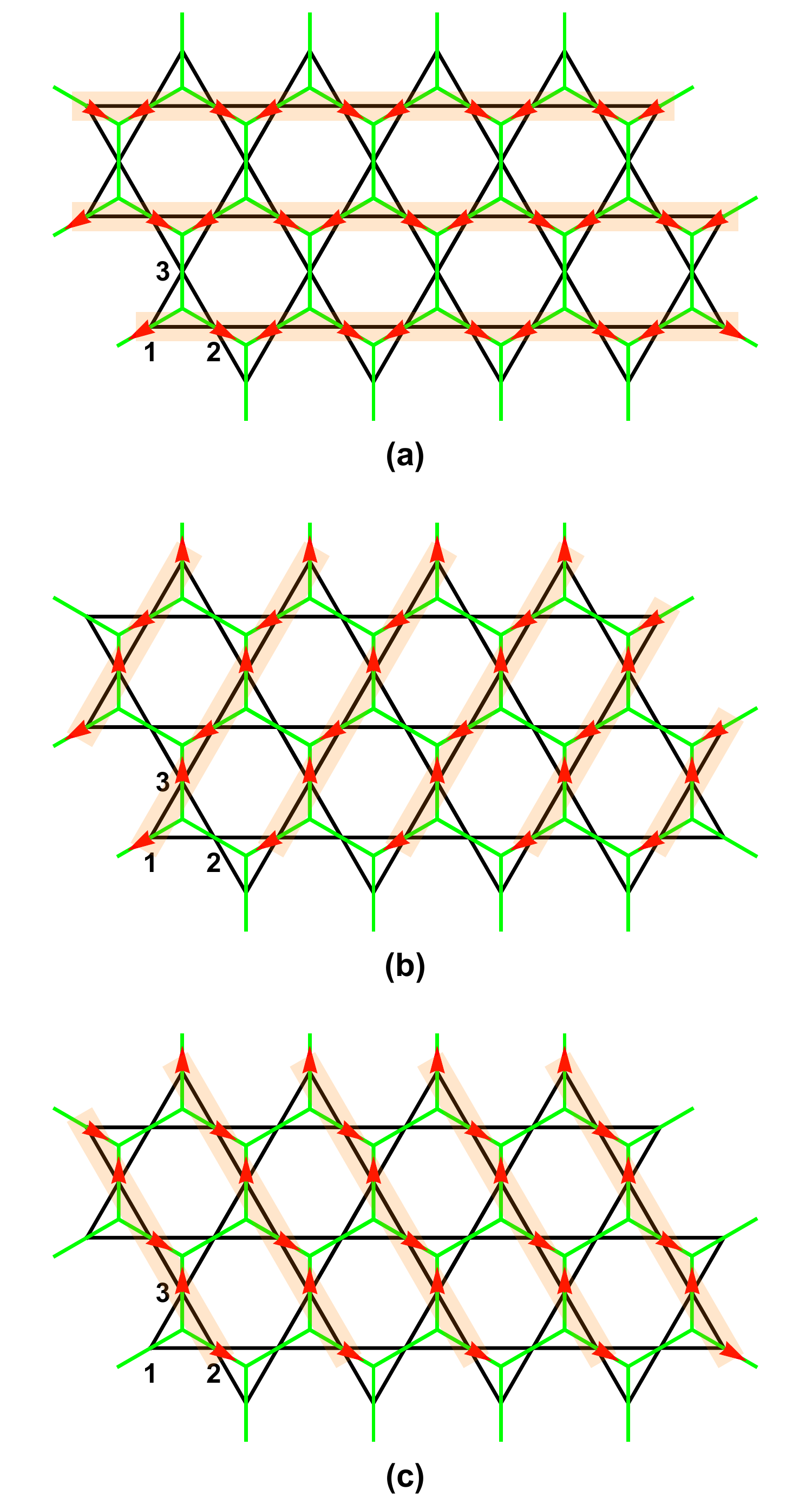}
	\caption{\label{dsf} Relation between 1-form charge (orange shaded ribbon) and the dynamical structure factor along specific momentum path. (a) $q_x=0$ path, (b) $q_{y}=-\frac{q_{x}}{\sqrt{3}}$ path, (c) $q_{y}=\frac{q_{x}}{\sqrt{3}}$. 1,2,3 label the three sublattice of the kagome lattice.}
\end{figure}

In this path $\mathbf{Q}=(0,q)$, denote $R_{i}=(x,y)$, then
\begin{equation}
\begin{aligned}
S^{z}_{-\mathbf{Q}}=&\frac{1}{L}\sum_{i}e^{i\,qy}\sum_{\alpha}e^{i\mathbf{Q}\cdot \rho_{\alpha}}S_{i,\alpha}^{z}\\
=&\frac{1}{L}\sum_{i}e^{i\,qy}[e^{-i\frac{q}{4\sqrt{3}}}(S^{z}_{i,1}+S^{z}_{i,2})+e^{-i\frac{q}{2\sqrt{3}}}S^{z}_{i,3}]\\
=&\frac{1}{L}\sum_{y}e^{i\,qy}\sum_{x}[e^{-i\frac{q}{4\sqrt{3}}}(S^{z}_{i,1}+S^{z}_{i,2})+e^{-i\frac{q}{2\sqrt{3}}}S^{z}_{i,3}],
\end{aligned}
\end{equation}
$\sum_{x}(S^{z}_{i,1}+S^{z}_{i,2})$ and $\sum_{x}S^{z}_{i,3}$ are both 1-form charges, this can be seen in Fig.\ref{dsf} (a). Hence $S^{z}_{-\mathbf{Q}}$ is conserved in the low energy theory, so there is no contribution to $S^{zz}(\mathbf{Q},\omega)$ from the ground state manifold. 

\subsection{$q_{y}=\pm\frac{q_{x}}{\sqrt{3}}$ path}
We first consider path $q_{y}=-\frac{q_{x}}{\sqrt{3}}$, where $\mathbf{Q}=(q,-\frac{1}{\sqrt{3}})$
\begin{equation}
\begin{aligned}
S^{z}_{-\mathbf{Q}}=&\frac{1}{L}\sum_{i}e^{i\,(qx-\frac{1}{\sqrt{3}}qy)}\sum_{\alpha}e^{i\mathbf{Q}\cdot \rho_{\alpha}}S_{i,\alpha}^{z}\\
=&\frac{1}{L}\sum_{i}e^{i\,(qx-\frac{1}{\sqrt{3}}qy)}[e^{-i\frac{q}{6}}(S^{z}_{i,1}+S^{z}_{i,3})+e^{-i\frac{q}{3}}S^{z}_{i,2}]\\
=&\frac{1}{L}\sum_{n}e^{i\,qn}\sum_{x-\frac{y}{\sqrt{3}}=n}[e^{-i\frac{q}{6}}(S^{z}_{i,1}+S^{z}_{i,3})+e^{-i\frac{q}{3}}S^{z}_{i,2}],
\end{aligned}
\end{equation}
$\sum_{x-\frac{y}{\sqrt{3}}=n}(\cdots)$ is summing over the shaded lines shown in Fig.\ref{dsf} (b). One can recognize that $\sum_{x-\frac{y}{\sqrt{3}}=n}(S^{z}_{i,1}+S^{z}_{i,3})$ and $\sum_{x-\frac{y}{\sqrt{3}}=n}S^{z}_{i,2}$ are also 1-form charges. So there will be no intensity in $S^{zz}(\mathbf{Q},\omega)$ in low energy. 

One can also find there is no intensity in $S^{zz}(\mathbf{Q},\omega)$ in low energy along path $q_{y}=\frac{q_{x}}{\sqrt{3}}$. The corresponding 1-form charges are shown in Fig.\ref{dsf} (c). This is actually the $\Gamma-M$ path (M point has momentum $(\pi,\frac{\pi}{\sqrt{3}})$). And this has been observed in our recent quantum Monte-Carlo study \cite{liu2023}.

\subsection{$q_{y}=0$ path}
$\Gamma-K$ path (K point has momentum $(\frac{4\pi}{3},0)$) belongs to this path.
\begin{equation}
\begin{aligned}
S^{z}_{-\mathbf{Q}}=&\frac{1}{L}\sum_{i}e^{i\,qx}\sum_{\alpha}e^{i\mathbf{Q}\cdot \rho_{\alpha}}S_{i,\alpha}^{z}\\
=&\frac{1}{L}\sum_{i}e^{i\,qx}[e^{-i\frac{q}{4}}S^{z}_{i,1}+e^{i\frac{q}{4}}S^{z}_{i,2}+S^{z}_{i,3}],
\end{aligned}
\end{equation}
this is not conserved unless $q=0$. When $\mathbf{Q}=0$, $S^{z}_{-\mathbf{Q}}=\frac{1}{L}\sum_{i,\alpha}S^{z}_{i,\alpha}$, and this is conserved in the original spin model eq.(\ref{ham}), so there is no intensity at $\mathbf{Q}=0$. 

At K point, 
\begin{equation}
\begin{aligned}
&e^{-i\frac{q}{4}}S^{z}_{i,1}+e^{i\frac{q}{4}}S^{z}_{i,2}+S^{z}_{i,3}\\
=&e^{-i\frac{\pi}{3}}S^{z}_{i,1}+e^{i\frac{\pi}{3}}S^{z}_{i,2}+S^{z}_{i,3},
\end{aligned}
\end{equation}
this is related to the order parameter of the PVBS phase \cite{Isakov2006prl,XFZhang2018prl}. So there is very large intensity in $S^{zz}(\mathbf{Q},\omega)$ in low energy at K point. Since the low energy excitations in the low energy manifold are created by the contractible loop operator, which can be seen as the local excitation, so the dispersions of these excitations are expected to be nearly flat. These expectations of the intensity along this path are also consistent with our recent numerical work \cite{liu2023}.

Here we see that the low energy part of the dynamical structure factor is constrained by the emergent 1-form U(1) symmetry in the low energy manifold. So the vanishing intensity along the high symmetry momentum paths can be viewed as the selection rule from the 1-form symmetry. As mentioned in last section, this 1-form symmetry is also exact in the sense that it is robust against any local UV perturbations which preserve translation symmetry. So this selection rule will also apply to a series of models which are obtained by perturbing around the classical part of the model (i.e. eq.(\ref{Isingm})).

\section{Discussion}\label{secdis}
\begin{figure}
	\centering
	\includegraphics[width=0.48\textwidth]{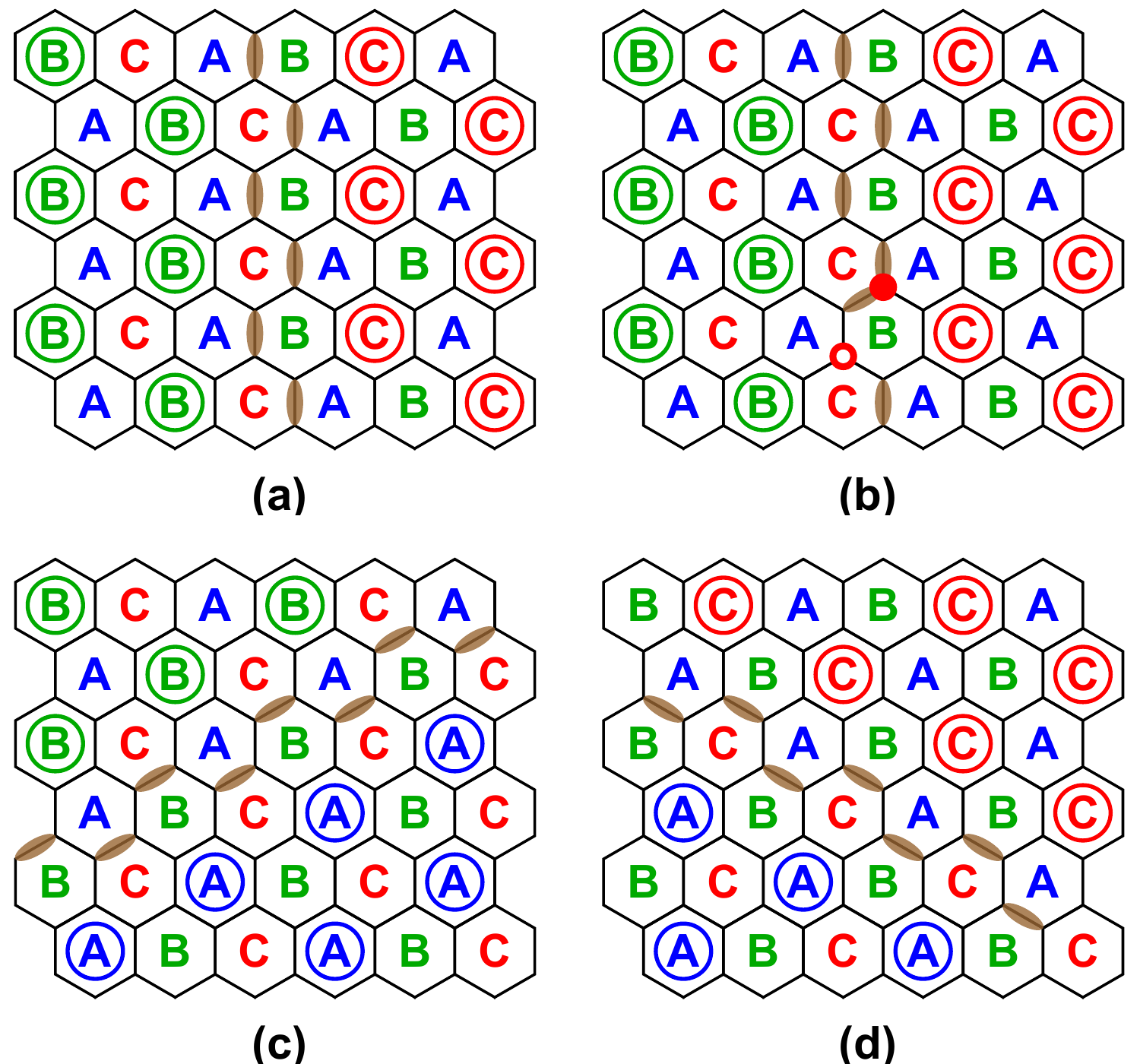}
	\caption{\label{dw} (a),(c),(d) Domain wall configurations in the threefold PVBS phase. (b) The charge excitations in the domain wall. The filled circle and the empty circle carry opposite charge. }
\end{figure}

Now we consider the excitations out of the low energy manifold. The simplest one is the charge excitation which breaks the Gauss's law. For example, $S^{+}S^{-}+h.c.$ can rotate the dimer around a site of honeycomb lattice, see Fig.\ref{dw} (b), this process can create a pair of opposite charges on the honeycomb lattice sites. Since the dimer is described by $E$ field, and the $a$ field can change the $E$ field configuration. The change of the charges will always accompanying by the change of the field, which is just the minimal coupling between the matter field and the gauge field. Thus, one can write down
\begin{equation}
S^{\pm}_{xy}=\rm{exp}\left[\pm i \eta_{x}\left(\textit{a}_{xy}+\Theta_{x}-\Theta_{y}\right)\right],
\end{equation}
where $e^{\pm i \Theta}$ is the ladder operator for the charge excitation, and it is canonical conjugated to the number operator $\mathcal{N}$ of the charge. Then the simplest theory \cite{XFZhang2018prl} for the matter-gauge coupling can be 
\begin{equation}\label{ahm}
\begin{aligned}
\mathcal{L}=&-\frac{1}{4e^2}|da|^2+\mathcal{L}_M+\sum_{i=1}^{2}[|D_{a}\varphi_i|^2+m^2|\varphi_i|^2+u|\varphi_i|^4]\\
&+w|\varphi_1|^2|\varphi_2|^2+\cdots,
\end{aligned}
\end{equation}
where $\mathcal{L}_M$ is the triple monopole term, $D_a$ is the covariant derivative. And $\varphi_i, i=1,2$ reflects the fact that there are two sublattices of the honeycomb lattice, thus the charge is two flavored. The first two terms are dual to the sine-Gordon theory eq.(\ref{sing}). And the mass term $m^2$ reflects that the charge excitation has a large gap in the PVBS phase. As we mentioned in Sec.\ref{gaugethe}, there should be also a term which breaks the charge conjugation symmetry, where we put this into the ellipsis. 

Actually, eq.(\ref{ahm}) is the abelian Higgs model with charge 3 monopole. The global symmetries of this model are the U(1) flavor symmetry, $Z_{2}$ exchange symmetry $\varphi_1 \leftrightarrow \varphi_2$, and the $Z_3$ topological symmetry which relates to the triple monopole \cite{Zohar2018prb,Sachdev1990prb}. The U(1) flavor symmetry and the $Z_{2}$ exchange symmetry can be extended to a O(2) symmetry. It is known that there are no nontrivial three dimensional symmetry protected topological phase in the system with these symmetries (O(2) and $Z_3$), this can be understood by decorated domain wall method \cite{Ashvin2013prx,XChen2014nc,CMJian2018prb}. Thus there is no mixed 't Hooft anomaly \cite{Metlitski2018prb} in eq.(\ref{ahm}).

It is worth noting that the abelian Higgs model with charge 2 monopole has mixed 't Hooft anomaly \cite{Tin2017prl,Zohar2018prb,Zohar2019}, if there is also charge conjugation symmetry, then the domain wall between the two fold degenerate PVBS configurations will also carries mixed anomaly \cite{Zohar2018prb,Zohar2019}. However, there is no anomaly in our case, so there is also no obvious constraint for the domain wall theory. Some domain wall configurations are shown in Fig.\ref{dw} (a),(c),(d), and there can be some charge excitations in the domain wall (see Fig.\ref{dw} (b)). Based on these analysis, we can only expect that the high energy part is some continuum in the dynamical structure factor. Further, these charge excitations will depend on the UV Hamiltonian, for example, the transverse field term $S^{x}$ will not conserve the dimer number, thus there may be no universal feature for the high energy spectrum.

\begin{acknowledgements}
We thank Liujun Zou, Shang-Qiang Ning and Salvatore Pace for helpful discussions. This work is supported by the International Postdoctoral Exchange Fellowship Program 2022 by the Office of China Postdoctoral Council: No.PC2022072 and the National Natural Science Foundation of China: No.12147172.
\end{acknowledgements}

\newpage
\bibliographystyle{apsrev4-1}
\bibliography{kagome_gauge}
\end{document}